\documentclass[letterpaper]{article} 
\usepackage{aaai24}  
\usepackage{times}  
\usepackage{helvet}  
\usepackage{courier}  
\usepackage[hyphens]{url}  
\usepackage{graphicx} 
\usepackage{natbib}  
\usepackage{caption} 
\usepackage{algorithm}
\usepackage{algorithmic}
\usepackage{multirow}
\usepackage{array}
\usepackage{booktabs}
\usepackage{adjustbox}
\usepackage{amssymb}
\usepackage{graphicx}
\usepackage{subcaption}
\usepackage{tikz}

\usepackage{newfloat}
\usepackage{listings}
\DeclareCaptionStyle{ruled}{labelfont=normalfont,labelsep=colon,strut=off} 
\floatstyle{ruled}
\newfloat{listing}{tb}{lst}{}
\floatname{listing}{Listing}
\title{SeGA: Preference-Aware Self-Contrastive Learning with Prompts for Anomalous User Detection on Twitter}
\author{
    Ying-Ying Chang, Wei-Yao Wang, Wen-Chih Peng
}
\affiliations{
    National Yang Ming Chiao Tung University, Hsinchu, Taiwan \\
    cindy88409.cs10@nycu.edu.tw, sf1638.cs05@nctu.edu.tw, wcpeng@cs.nycu.edu.tw
}

\begin{document}

\maketitle

\begin{abstract}
In the dynamic and rapidly evolving world of social media, detecting anomalous users has become a crucial task to address malicious activities such as misinformation and cyberbullying.
As the increasing number of anomalous users improves the ability to mimic normal users and evade detection, existing methods only focusing on bot detection are ineffective in terms of capturing subtle distinctions between users.
To address these challenges, we proposed SeGA, preference-aware self-contrastive learning for anomalous user detection, which leverages heterogeneous entities and their relations in the Twittersphere to detect anomalous users with different malicious strategies.
SeGA utilizes the knowledge of large language models to summarize user preferences via posts.
In addition, integrating user preferences with prompts as pseudo-labels for preference-aware self-contrastive learning enables the model to learn multifaceted aspects for describing the behaviors of users.
Extensive experiments on the proposed TwBNT benchmark demonstrate that SeGA significantly outperforms the state-of-the-art methods (+3.5\% $\sim$ 27.6\%) and empirically validate the effectiveness of the model design and pre-training strategies.
Our code and data are publicly available at https://github.com/ying0409/SeGA.





\end{abstract}

\vspace{-8pt}
\section{Introduction}

The exploration of anomalous user detection has broad applicability across various domains.
Whether it involves developing strategies for network security \cite{DBLP:journals/access/BilotMAZ23}, financial transactions \cite{DBLP:conf/ijcai/ChaiY0PXCJ22}, or social media analytics \cite{DBLP:conf/sigir/AgarwalSSR22,DBLP:conf/aaai/WangP22}, these scenarios can be effectively framed as anomalous user detection systems characterized by intricate relations between users.
Twitter serves as one of the most widely-used social media platforms; however, the widespread features in recent years have also led to the rapid growth of anomalous users.
For instance, abnormal users often initiate campaigns to pursue malicious goals, which violates the principles of healthy online discussions on social media platforms \cite{DBLP:journals/snam/AlievaMC22}.
One of the most common types of anomalous users is bots, identified at 9\% to 15\% of active users \cite{DBLP:conf/icwsm/VarolFDMF17}.
In late 2017, the US Congress disclosed a list of 2.7k Twitter accounts that were identified as Russian trolls \cite{DBLP:conf/www/ZannettouCCSSB19}, which is an emerging type of anomalous user.
As the variety of anomalous user types increases, and their potential negative impact on social media becomes more pronounced, it is crucial to emphasize the significance of advancing anomaly detection methods capable of identifying not only bots but also trolls.

Previous works primarily focused on identifying bots and achieved effectiveness by leveraging the topology with graph neural networks (GNNs) \cite{DBLP:conf/iclr/KipfW17, DBLP:conf/iclr/VelickovicCCRLB18} and heterogeneous information networks (HINs) \cite{DBLP:conf/asunam/FengWWL21,DBLP:conf/aaai/FengTLL22}.
However, as the diversity of anomalous users and the evolutionary strategies for malicious activities continue to expand, such as spreading fake news through lists\footnote{A list is a curated selection of Twitter accounts for organizing.}, existing methods lack the ability to effectively distinguish various types of anomalous users like both bots and trolls.
The main difference in detecting trolls and bots is that the former is controlled by humans, resulting in similar behavior to normal users compared with the latter.

\begin{figure}[]
  \centering
  \includegraphics[width=0.90\linewidth]{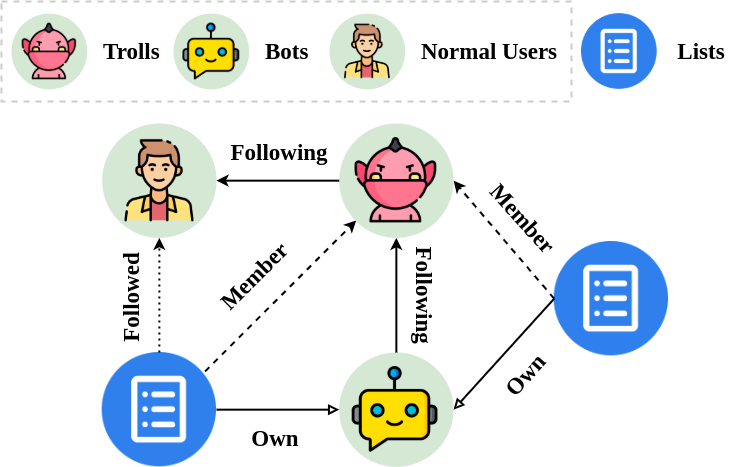}
  \caption{An example of anomalous user behavior in social media, where the edges connecting entities represent diverse relationships.}
  \label{fig:anomaly_user_detection}
\end{figure}

Figure \ref{fig:anomaly_user_detection} illustrates an example of anomalous user behavior with other entities on social media, where anomaly users are able to not only interact with other users, but also own lists and add other anomalous users as members to carry out malicious activities such as spreading fake news.
For example, a bot might create a list and add trolls as members.
If a normal user follows this list, they may receive misinformation that is spread by anomalous users within the list.
Therefore, a challenging problem arises with this problem: \textit{How to capture the subtle difference between users when troll users may behave similarly to normal users?}

To tackle these issues, we propose a novel framework SeGA, a preference-aware self-contrastive learning approach for anomalous user detection on Twitter, which encodes the heterogeneous relations of various entities on Twitter with the heterogeneous encoder.
We introduce self-contrastive learning with pseudo-labels to discern subtle differences between users based on user preferences via the corresponding posts.
Specifically, we construct an HIN that incorporates various edge types between node types including users and lists to model users with diverse activities.
In order to learn about discrepancies between users, the pre-training strategy incorporates the knowledge of large language models (LLMs) to capture user-preferred topics and emotions for preference-aware self-contrastive learning with prompts.
To evaluate the effectiveness of our proposed method, we propose a new anomalous user detection benchmark, TwBNT, which demonstrates a significant improvement of at least 3.5\% compared with the state-of-the-art baselines in terms of F1 score.

Our main contributions are summarized as follows:
\begin{itemize}
    \item We propose SeGA to address the challenging but emerging anomalous user detection task on Twitter. To the best of our knowledge, this is the first work that jointly distinguishes normal, troll, and bot users on social media.
    \item We introduce preference-aware self-contrastive learning to learn user behaviors via the corresponding posts. In addition, we incorporate prompt templates with user preferences as pseudo-labels to capture the user-preferred topics and emotions.
    \item We collected a large-scale Twitter dataset named TwBNT for anomalous user detection including normal users, troll users, and bots. Extensive experiments were conducted to evaluate the performance of our proposed model, which demonstrated superior improvement of between 3.5\% and 27.6\%. 
\end{itemize}

\section{Preliminaries}
\subsection{Related Work}

\subsubsection{Anomalous User Detection on Twitter}
Early Twitter anomalous user detection models focused on detecting bot accounts with user features or tweets \cite{DBLP:journals/isci/MillerDDHW14, DBLP:journals/expert/CresciPPST16}.
With the advent of graph neural networks, an increasing number of graph-based bot detectors have been proposed by representing users and their interactions as a social graph, and utilizing aggregation techniques to gather information from neighboring nodes.
For example, GCN \cite{DBLP:conf/iclr/KipfW17} aggregates features equally from neighboring users to learn representations, while GAT \cite{DBLP:conf/iclr/VelickovicCCRLB18} models user influence using attention mechanisms.
On the other hand, heterogeneous graphs are also utilized for bot detection due to their effectiveness in representing social networks with diverse node and edge types.
\citet{DBLP:conf/kdd/LvDLCFHZJDT21} adopted different strategies to enhance GAT on the heterogeneous graph, and \citet{DBLP:conf/aaai/FengTLL22} proposed relational graph transformers to model heterogeneous relations and influence heterogeneity between users.
Nonetheless, previous methods failed to detect another important facet of anomalous users: trolls, which react similarly to normal users to avoid detection.
Therefore, we introduce preference-aware self-contrastive learning to differentiate user preferences from posts with pseudo-labels.

\subsubsection{Self-Supervised Learning}
In recent years, self-supervised learning (SSL) has been proven to be a powerful and effective approach in various domains for learning contextualized representations via pretext tasks, such as natural language processing \cite{DBLP:conf/emnlp/GaoYC21} and computer vision \cite{DBLP:conf/kdd/LvDLCFHZJDT21, DBLP:conf/nips/BardesPL22}.
Predictive learning serving as a branch of SSL has also been applied for bot detection with promising results.
\citet{DBLP:conf/cikm/FengWWLL21} incorporated follower count as a self-supervised learning signal to enhance the model's performance in bot detection.
However, relying on a single feature as a self-supervised indicator overlooks the diversity among anomalous users and falls short of adequately representing different users.
For instance, troll users are able to construct artificial follower counts followed by other troll users.
Therefore, we adopt user preferences acquired by LLMs with the corresponding posts to summarize multifaceted behaviors (topics and emotions in this paper) to describe users for self-contrastive learning.

\subsection{Problem Formulation}

We denote $G = \{V, E, A, R, \phi, \psi\}$ as the input heterogeneous information network (HIN) with different types of nodes $V$ and edges $E$, and it is associated with a node mapping function $\phi: V \rightarrow A$ and an edge mapping function $\psi: E \rightarrow R$, where $A$ and $R$ denote the sets of all node types and edge types.
The nodes in HIN are categorized into either user type $u$ or list type $l$ as $v^A_i \in V$, where $A = \{u,l\}$ is the node type of node $i$ representing user and list, respectively.
The edges are denoted as $(v^A_i,r,v^A_j) \in E$, where $r \in R$ represents the relation between node $i$ and node $j$.

For each node $i$, node features are categorized into three types: indicator features $C_i = \{c_{i_1}, c_{i_2},..., c_{i_k}\}$ indicating if matching the indication, numerical features (e.g., number of posts) $N_i = \{n_{i_1}, n_{i_2},..., n_{i_m}\}$ and textual features $T_i = \{des_{i}, twe_{i}\}$, where $des_{i} = \{w^{des}_{i_1}, w^{des}_{i_2},...,w^{des}_{i_s}\}$ denotes the description with $s$ words for a user or a list, and $twe_{i} = \{{twe}_{i_j}\}_{j=1}^{q}$ denotes $q$ tweets posted from a user or contained in a list, while each tweet $j$ contains $L$ words denoted as ${twe}_{i_j} = \{w^{twe}_{i_{j_1}}, w^{twe}_{i_{j_2}}, \cdots , w^{twe}_{i_{j_L}}\}$.
We formulate anomalous user detection as a multi-class classification problem, where users are classified into three classes based on past activities and relations with society.
Given a learned HIN $G$, we aim to learn an anomalous user detection function $f(G) \rightarrow \hat{y}$ to predict whether the user $v^u_i$ is a troll, bot, or normal user.

\section{The Proposed Method}

\begin{figure*}[t]
  \centering
  \includegraphics[width=0.90\linewidth]{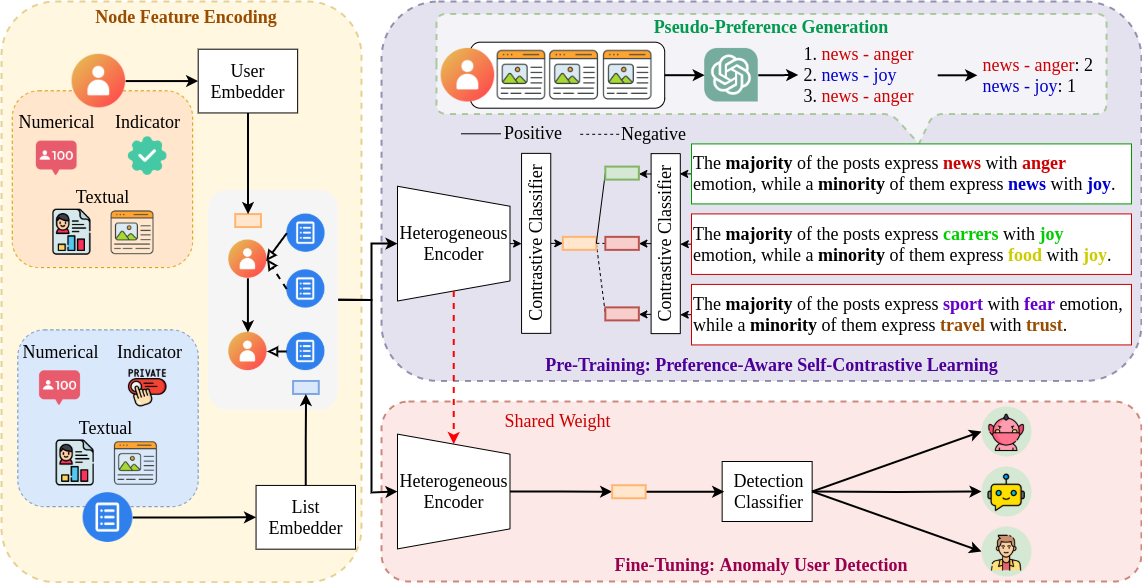}
  \caption{Overview of our proposed framework, SeGA, consisting of three stages: 1) Node feature encoding to initialize node embeddings from heterogeneous information; 2) The pre-training stage to pre-train the encoder with pseudo-aware self-contrastive learning to capture the differences in user preferences; and 3) The fine-tuning stage to classify anomalous users. }
  \label{fig:framework}
\end{figure*}


The overview of SeGA is shown in Figure \ref{fig:framework}, which consists of three stages: node feature encoding, the pre-training stage, and the fine-tuning stage.
SeGA leverages users and lists as nodes within an HIN to capture diverse relationships among these entities.
We first encode three types of node features (i.e., indicator, numerical, and textual) to encode representations for each node type.
In the pre-training stage, the encoder is pre-trained with preference-aware self-contrastive learning to learn the user preferences summarized by LLMs from posts.
In the fine-tuning stage, we fine-tune the pre-trained model to classify anomalous users.

\subsection{Model Architecture}
\subsubsection{Node Feature Encoding}
Similar to the process of feature encoding as \cite{DBLP:conf/aaai/FengTLL22}, we first concatenate each indicator feature for each node $i$ with node type $A$:
\begin{equation}
\small
x^{A_{ind}}_i = c_{i_1} \oplus c_{i_2} \oplus \cdots \oplus c_{i_k},
\end{equation}
where $\oplus$ is the concatenation operator, and $x^{A_{ind}}_i \in \mathbb{R}^{k}$ is the indicator embedding concatenated by $k$ features.

Similarly, the numerical features are concatenated as:
\begin{equation}
\small
x^{A_{num}}_i = n_{i_1} \oplus n_{i_2} \oplus ... \oplus n_{i_m},
\end{equation}
where $x^{A_{num}}_i \in \mathbb{R}^{m}$ is the numerical embedding for node $i$ concatenated by $m$ features.

To encode textual features, we applied pre-trained RoBERTa \cite{DBLP:journals/corr/abs-1907-11692} for encoding with $s$ words:
\begin{equation}
\small
x^{A_{des}}_{i} = RoBERTa(\{w^{des}_{i_j}\}^s_{j=1}),
\end{equation}
where $x^{A_{des}}_{i} \in \mathbb{R}^{d_{des}}$ denotes the description embedding.
Likewise, we obtain the tweet embedding $x^{A_{twe}}_{i_j} \in \mathbb{R}^{d_{twe}}$ by averaging all embeddings encoded from each tweet with pre-trained RoBERTa:
\begin{equation}
\small
x^{A_{twe}}_{i_j} = RoBERTa(\{w^{twe}_{i_{j_y}}\}^L_{y=1}),
\end{equation}
\begin{equation}
\small
x^{A_{twe}}_{i} = AVG(x^{A_{twe}}_{i_1}, x^{A_{twe}}_{i_2}, \cdots, x^{A_{twe}}_{i_q}).
\end{equation}
Afterwards, we transform them separately:
\begin{equation}
\small
x'^{A_{F}}_{i} = \sigma(W_x \cdot x^{A_{F}}_{i}+b_x), F \in \{ind, num, des, twe\}
\end{equation}
where $x'^{A_{F}}_{i} \in \mathbb{R}^{d_h}$, $W_x$, $b_x$ are trainable parameters, and $\sigma$ is denoted as LeakyReLU.

For each user or list node $i$, we obtain the user/list embeddings $x^A_i \in \mathbb{R}^{4*d_h}$ by concatenating the indicator $x'^{A_{ind}}_i$, numerical $x'^{A_{num}}_i$, description $x'^{A_{des}}_i$ and tweet $x'^{A_{twe}}_i$ embeddings:
\begin{equation}
\small
    x^A_i = x'^{A_{ind}}_i \oplus x'^{A_{num}}_{i} \oplus x'^{A_{des}}_{i} \oplus x'^{A_{twe}}_{i}.
\end{equation}
We then transform the user/list embeddings $x^A_i$ as the initial node embedding $z^{A,(0)}_i \in \mathbb{R}^{4*d_h}$ for the heterogeneous encoder:
\begin{equation}
\small
    z^{A,(0)}_i = \sigma(W_I \cdot x^A_i + b_I),
\end{equation}
where $W_I$ and $b_I$ are trainable parameters.

\subsubsection{Heterogeneous Encoder}
To model the various entities and their diverse relation with different importance to enrich the embeddings of users, we applied a relational graph transformer (RGT) following \cite{DBLP:conf/aaai/FengTLL22} and an MLP with activation function as the heterogeneous encoder:
\begin{equation}
\small
    z^{u, (g)}_i, z^{l, (g)}_i = RGT^{(g)}(z^{u, (g-1)}_i, z^{l, (g-1)}_i),
\end{equation}
\begin{equation}
\small
    z^u_i = \sigma(W_z \cdot z^{u,(g)}_i + b_z),
\end{equation}
where $z^{u,(g)}_i, z^{l,(g)}_i \in \mathbb{R}^{d_{out}}$ is the $i$-th node embedding learned from the $g$-th layer, $z^u_i \in \mathbb{R}^{d_{u}}$ is the $i$-th user representations, and $W_z$ and $b_z$ are trainable parameters.
We note that as the task is to classify user categories, we adopt aggregated user embeddings $z^u_i$ for preference learning and anomaly detection.

\subsection{Pre-Training Stage: Learning User Preferences via Posts}
User preferences represent the behavior of an individual as they showcase a person's choices in various aspects, which is beneficial for detecting users' behavior patterns.
Therefore, we introduce the pseudo-preference generation, which is summarized from LLMs based on users' historical posts.
In order to describe and learn user preferences, we designed a prompt template to represent the majority and minority topic-emotion pairs of user posts for preference-aware self-contrastive learning.

\subsubsection{Pseudo-Preference Generation}
Given the success paradigms of leveraging LLMs in natural language applications \cite{DBLP:conf/acl/WuZH23}, we incorporate the powerful knowledge of LLMs to retrieve user preferences from posts.
Specifically, we opt for the preferred topic and emotion to represent the preference of each user from the corresponding posts since anomalous users may exploit them to achieve malicious intentions \cite{DBLP:journals/corr/abs-1910-01340,DBLP:conf/icwsm/Balasubramanian22}.
The 10 recent posts of user $i$ are used as the prompt for LLM to generate the topic $t$ and the emotion $e$ used in each tweet $j$:
\begin{equation}
\small
    \{(t^u_{i_j}, e^u_{i_j})_{j=1}^{10}\} = LLM(P(twe^u_{i_1} \oplus \cdots \oplus twe^u_{i_{10}})),
\end{equation}
where $P()$ is an instruction prompt to acquire the user preference from LLMs described in Appendix A.1, and $LLM$ is ChatGPT \cite{DBLP:journals/corr/abs-2304-01852} in this paper.
The topics are derived from 16 categories based on Twitter topics \cite{twitter_topic}, and the 8 emotions are based on Plutchik's emotions \cite{DBLP:journals/corr/abs-1910-01340}.

\subsubsection{Preference-Aware Self-Contrastive Learning}
After obtaining the topic-emotion pairs for each user, we aim to pre-train the model with this pseudo-information to describe user preferences.
An intuitive approach is to predict all topics and emotions of each post by a user, which can be formulated as a multi-label classification task.
However, this fails to effectively capture the user-preferred topics and emotions since the model treats all labels equally without considering their relative importance.
For example, a user may post mainly on a news topic with anger and seldom post on a news topic with joy.

To tackle this issue, we propose a preference-aware self-contrastive learning approach that incorporates preferences with prompts for learning enhancement.
We define the preference of a user as the integration of the most frequently used topic-emotion pair $(t^u_{i_{max}}, e^u_{i_{max}})$ and the least frequently used pair $(t^u_{i_{min}}, e^u_{i_{min}})$ to reflect the user's interest and emotional behavior.
We then generate the pseudo-label $p^u_i$ for each user $i$ by leveraging the designed prompt template $PT()$ filled with the most and the least frequently used topic emotion pairs as:
\begin{equation}
\small
    p^u_{i} = PT((t^u_{i_{max}}, e^u_{i_{max}}),(t^u_{i_{min}}, e^u_{i_{min}})).
\end{equation}
The prompt template is designed as: \textit{``The majority of the posts express $t^u_{i_{max}}$ with $e^u_{i_{max}}$ emotion, while a minority of them express $t^u_{i_{min}}$ with $e^u_{i_{min}}$."}.
Experiments with different templates are discussed in Section \ref{prompt-variant}.

Afterwards, pseudo-labels of user $i$ are encoded with the prompt encoder using SimCSE RoBERTa \cite{DBLP:conf/emnlp/GaoYC21}, which learns sentence embeddings from contrastive learning:
\begin{equation}
\label{prompt-encoder}
\small
    p'^u_i = SimCSE(p^u_i),
\end{equation}
where $p'^u_i \in \mathbb{R}^{d_p}$ is the pseudo-label embedding of user $i$.
Then, we transform the user embedding $z^u_i$ and pseudo-label embedding $p'^u_i$ with the contrastive classifier:
\begin{equation}
\small
    \tilde{z}^u_i = W_{\tilde{z}} \cdot z^u_i + b_{\tilde{z}},
\end{equation}
\begin{equation}
\small
    \tilde{p}^u_i = W_{\tilde{p}} \cdot p'^u_i + b_{\tilde{p}},
\end{equation}
where $\tilde{z}^u_i \in \mathbb{R}^{d_{a}}$ denotes the anchor user embedding for user $i$ and $\tilde{p}^u_i \in \mathbb{R}^{d_{a}}$ denotes the corresponding positive sample embedding, $W_{\tilde{z}}$, $b_{\tilde{z}}$, $W_{\tilde{p}}$ and $b_{\tilde{p}}$ are trainable parameters.

To compute the contrastive loss $L_{pre}$, we define the embedding $\tilde{p}^u_i$ as the positive pair of the anchor user's embedding $\tilde{z}^u_i$, and embeddings transformed from other pseudo-labels are considered as the negative pairs of the anchor user:
\begin{equation}
\small
    L_{pre} = -\sum_{i \in U} log\frac{exp(sim(\tilde{z}^u_i \cdot \tilde{p}^u_i)/\tau)}{\sum_{j \in S(i)}exp(sim(\tilde{z}^u_i \cdot \tilde{p}^u_j)/\tau)},
\end{equation}
where $U$ is the set of indices of all user nodes, $S(i)$ is the set of a positive pair, and negative pairs are randomly sampled from all negative pairs.
When minimizing the contrastive loss, the user embeddings with the same pseudo-label tend to be closer together, while simultaneously encouraging the encoder to learn user-preferred topics and emotions.

This generic pre-training objective with pseudo-labels and the prompt template reinforces the model to learn user behavior with at least two advantages: 
\begin{itemize}
    \item User preferences can also be utilized in other social media-related tasks, such as user recommendation \cite{DBLP:conf/www/WangXZWS22} and community detection \cite{DBLP:conf/cikm/WuZYLL21}, since posts play a crucial role in user interactions within social media.
    \item Incorporating user post preference enables an effective representation of users, as posting content is the primary avenue through which individuals pursue their objectives and express themselves on social media.
\end{itemize}

\subsection{Fine-Tuning Stage: Anomalous User Detection}
To achieve the goal of anomalous user detection, the user embedding $z^u_i$ from the pre-trained model is used in the detection classifier with a softmax layer to predict the class $\hat{y}$ for each user $i$:
\begin{equation}
\small
    \hat{y}_i = softmax(W_{y} \cdot z^u_i + b_{y}),
\end{equation}
where $W_{y}$, $b_{y}$ are trainable parameters.

Finally, we jointly fine-tune the pre-trained embedders, pre-trained encoder, and the detection classifier for anomalous user detection.
The fine-tuning process incorporates both the cross-entropy loss and an L2 regularization term as follows:
\begin{equation}
\small
    L_{fine} = -\sum_{i \in U} [y_i \log(\hat{y}_i)] + \lambda \sum_{\omega \in \theta} \omega^2,
\end{equation}
where $U$ is the set of indices of all user nodes, $y_i$ represents the ground-truth label of user $i$, $\lambda$ is a hyper-parameter, and $\theta$ encompasses all the trainable parameters.

\section{Experiments}

\subsection{Experiment Settings}
\subsubsection{Proposed Benchmark: TwBNT Dataset}
Since there is no public dataset consisting of trolls, bots, and normal users, we proposed a new dataset, TwBNT, by extending the bot detection benchmark Twibot-22 \cite{DBLP:conf/nips/FengTWWCZZZLYFZ22} with automatic troll annotations.
Twibot-22 is a bot detection dataset that provides graph structures with various entities and relations within the Twitter network, which evaluates graph-based approaches to bot detection.
Nonetheless, Twibot-22 annotates users as either bots or humans, but excludes the critical category for trolls.
To this end, we sample users from Twibot-22 using a breadth-first search algorithm for user collection following \cite{DBLP:conf/cikm/FengWWLLL21} to ensure the sampled users include different types of bots, trolls, and normal users.
Then, the list nodes connected to the sampled users are employed to construct the TwBNT dataset.
Since trolls are controlled by real human users, we automatically identify them by obtaining a troll score $scr \in [0,1]$ for each user labeled in Twibot-22 with the widely recognized platform Bot Sentinel\footnote{https://botsentinel.com/}.
Users with a score greater than the threshold value $scr = 0.5$ are labeled as trolls, while others are labeled as normal users following the spreading intent detection \cite{DBLP:conf/www/ZhouSP0Z22}.
Table \ref{tab:dataset} summarizes the statistics of our collected datasets.

\subsubsection{Baselines}
Due to the lack of baselines for the proposed anomaly user detection tasks that also distinguish troll users, we compared the proposed model with 6 baselines of bot detection methods to verify its effectiveness: GCN \cite{DBLP:conf/iclr/KipfW17}, GAT \cite{DBLP:conf/iclr/VelickovicCCRLB18}, SimpleHGN \cite{DBLP:conf/kdd/LvDLCFHZJDT21}, SATAR \cite{DBLP:conf/cikm/FengWWLL21}, BotRGCN \cite{DBLP:conf/asunam/FengWWL21}, and RGT \cite{DBLP:conf/aaai/FengTLL22}.
More details of baselines are given in Appendix B.

\subsubsection{Implementation Details}
The numbers of indicator features $k$ for the user and list node are 3 and 1, and the numbers of numerical features $m$ for the user and list nodes are 5 and 4.
Detailed features are described in Appendix A.2.
The numerical features $N$ are applied with z-score normalization.
The dimensions of $d_{des}$, $d_{twe}$ and $d_p$ are 768, the dimension of $d_h$ is 32, the dimension of $d_{out}$ is 128, and the dimensions of $d_u$ and $d_a$ are 64.
The number of layers $g$ of the relational graph transformer is 2.
Following \cite{DBLP:conf/aaai/FengTLL22}, the max number of tweets $q$ for each user and list is set to 20 for representing the recent 20 tweets from each user and list.
The max lengths for description $s$ and each tweet $L$ are set to 50 words, and padding with zeros is applied if the length is insufficient.
In the pre-training stage, we ask ChatGPT to classify each tweet into 16 topics and 8 emotions by providing the instructions depicted in Appendix A.2.
As there remain some results that do not belong to these categories, we treat them as ``others".
We randomly sample 100 prompts with other pseudo-labels as negative samples and set the temperature $\tau$ as 0.1 for computing the pre-training loss.
In the fine-tuning stage, $\lambda$ is set to $3 \times 10^{-5}$.
The training epochs of the pre-training and fine-tuning stages are set to 100 and 150, respectively.
The dropout rate is set to 0.3.
We employ the AdamW optimizer \cite{DBLP:conf/iclr/LoshchilovH19} using the learning rate of 0.001 and the batch size is set to 2048.
All experiments were conducted with a Nvidia RTX A5000 GPU and all parameters were tuned based on the validation set.
The analysis of the hyper-parameter is discussed in Appendix C.1.

\begin{table}
    \caption{Statistics of the proposed benchmark TwBNT.}
    \label{tab:dataset}
\begin{center}
\begin{tabular}{c|cc|c}
    \toprule
    Types & Item & Count & Total\\
    \midrule
    \multirow{2}{*}{2 node types ($A$)} & \# user & 100,001 & \multirow{2}{*}{120,789}\\ & \# list & 20,788 \\
    \midrule
    \multirow{5}{*}{5 edge types ($R$)}  & \# following & 751,927 & \multirow{5}{*}{1,312,929}\\ & \# followers & 148,250 \\ & \# membership & 365,593 \\ & \# followed & 39,258 \\ & \# own & 7,901 \\
    \midrule
    \multirow{3}{*}{3 classes} & \# troll & 896 & \multirow{3}{*}{100,001}\\ & \# bot & 5,047 \\ & \# normal & 94,058 \\ 
    \midrule
    \multirow{3}{*}{3 splits} & \# train & 70,000 & \multirow{3}{*}{100,001}\\ & \# valid & 20,000 \\ & \# test & 10,001 \\ 
    \bottomrule
\end{tabular}
\end{center}
\end{table}

\begin{table*}[!]
    \centering
    \small
    \caption{Performance of all baseline methods and our proposed SeGA. All methods are evaluated by precision, recall, and Macro F1. The best results are highlighted in boldface, and the second-best results are underlined. The strategies employed by each method are marked as \checkmark.}
    \label{tab:compare_performance}
    \newcommand{\specialcell}[2][c]{
  \begin{tabular}[#1]{@{}c@{}}#2\end{tabular}}

\begin{adjustbox}{center}
\begin{tabular}{c|cccc|cc|c}
\toprule
Methods & \specialcell[c]{Attention \\Mechanism} & \specialcell[c]{Edge \\Heterogeneity} & \specialcell[c]{Node \\Heterogeneity} & \specialcell[c]{Self\\ Supervision} & Precision & Recall & F1 \\
\midrule
GCN & & & & & 67.43$\pm$1.24 & 42.76$\pm$1.14 & 47.57$\pm$1.54\\
GAT & \checkmark & & & & 66.40$\pm$2.19 & 45.54$\pm$0.84 & 50.75$\pm$0.66\\
SimpleHGN & \checkmark & \checkmark & & & 79.84$\pm$2.54 & 45.68$\pm$1.19 & 52.22$\pm$1.54 \\
SATAR & \checkmark & \checkmark & & \checkmark & 55.31$\pm$7.69 & 44.77$\pm$2.35 & 46.98$\pm$1.06 \\
BotRGCN & & \checkmark & & & 65.85$\pm$12.55 & 43.51$\pm$5.20 & 47.94$\pm$7.20 \\
RGT & \checkmark & \checkmark & & & 75.55$\pm$1.27 & 52.10$\pm$1.25 & \underline{58.61$\pm$0.92}\\
\midrule
\midrule
SeGA (Ours) & \checkmark & \checkmark & \checkmark & \checkmark & 68.13$\pm$0.97 & 56.58$\pm$1.50 & \textbf{60.69$\pm$0.72} \\
\bottomrule
\end{tabular}
\end{adjustbox}
\end{table*}

\subsubsection{Evaluation Metrics}
As the numbers of troll and bot users are more imbalanced than normal users, as shown in Table \ref{tab:dataset}, macro F1 scores are utilized as principal indicators to evaluate the overall performance of the model.
Additionally, precision and recall are employed to evaluate the results of anomaly user classification.
We evaluate all baselines and our model by computing the mean and standard deviation for the results obtained with 3 different random seeds.

\subsection{Quantitative Performance}
Table \ref{tab:compare_performance} summarizes the overall performance of anomalous user detection methods, demonstrating that our proposed model surpasses all baselines, which are extended to classify trolls.
Quantitatively, SeGA achieves at least a 3.5\% improvement in F1-score compared with the best-performing baseline.
We summarize the observations as follows:

Compared with the homogeneous graph-based methods, GCN and GAT, SeGA achieves a significant improvement on Macro F1 with 27.6\% and 19.6\%, demonstrating the capability of heterogeneous graphs to classify various anomalous and normal users.
Additionally, SeGA outperforms other heterogeneous graph-based methods such as SimpleHGN, BotRGCN, and RGT, ranging from 3.5\% to 26.6\%, which is attributed to considering diverse entities not only from edges but also from nodes.
This also highlights the capability of leveraging topic-emotion pairs via posts with prompts as pseudo-labels to model user preferences.

Although SATAR also adopts the follower count as a self-supervised objective, we can observe that the performance of SATAR deteriorates substantially since troll users can manipulate the follower count followed by other troll users to act as normal users.
The comparison of SeGA and SATAR reveals the importance of considering user preferences for distinguishing anomalous users, where the user preference not only behaves as a task-agnostic objective but also describes the user profile based on the corresponding posts.

\begin{figure}
  \centering
  \includegraphics[width=\linewidth]{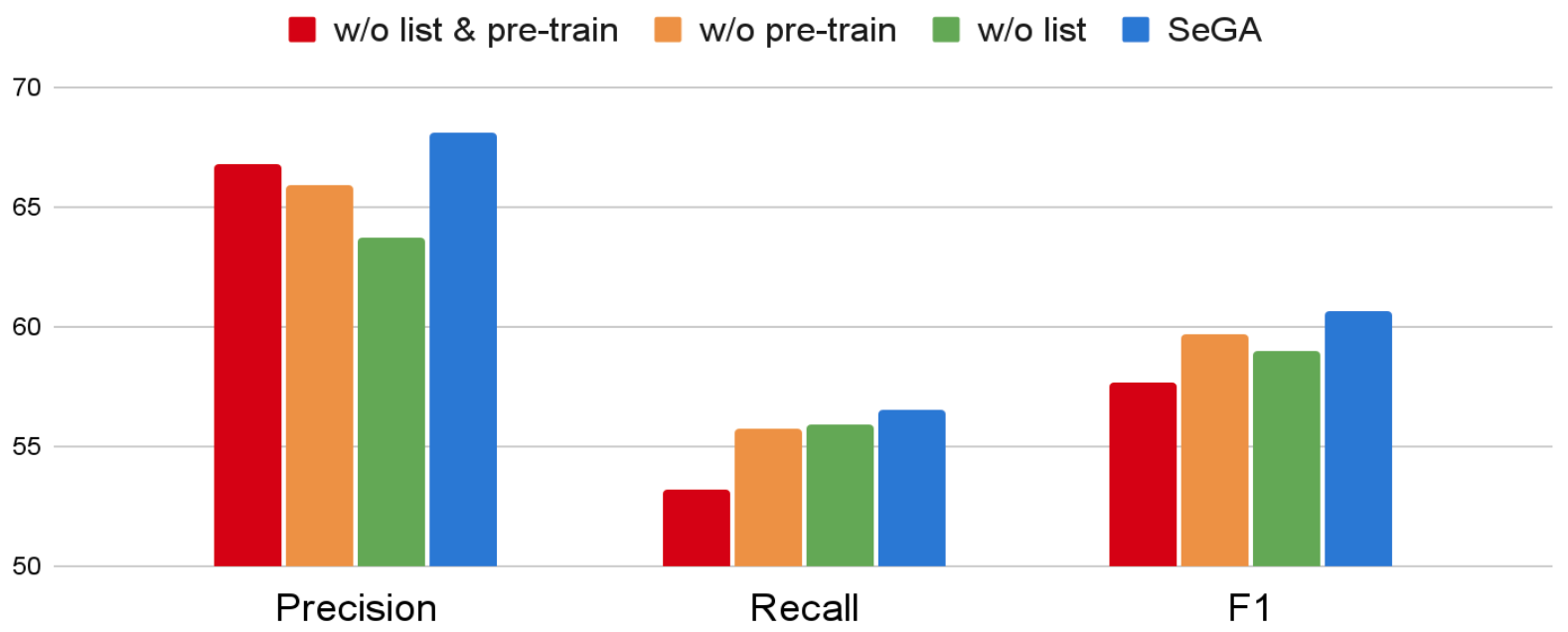}
  \caption{Ablation study including removing the list nodes and the pre-training stage from SeGA.}
  \label{fig:ablation}
\end{figure}

\subsection{Ablation Study}
To analyze the contributions of different designs of SeGA, we conducted an ablation study by removing different node types and the pre-training stage as shown in Figure \ref{fig:ablation}.
As expected, removing both types of information (w/o list \& pre-train) significantly hinders the performance.
Moreover, we can observe that the F1 score decreases when removing list information (w/o list) for constructing the graph, which signifies that considering user relations with lists helps in detecting anomalous users with various malicious activities.
The deleterious performance of removing the pre-training stage (w/o pre-train) suggests the advantage of learning user preferences as the pre-training objective.

\begin{table}
    \caption{Performance of SeGA with different pre-training settings, including modifications in the prompt encoder (A1), prompt design (A2), and pre-training task (A3).}
    \label{tab:prompt}
    \begin{adjustbox}{center}
\begin{tabular}{cc|cc|c}
    \toprule
     & Methods & Precision & Recall & F1 \\
    \midrule
    \multirow{1}{*}{A1} & RoBERTa & 68.32 & 55.39 & 59.84 \\
    \midrule
    \multirow{4}{*}{A2}
    & Short & 65.86 & 54.16 & 58.26 \\
    & Topic & 66.03 & 55.53 & 59.27 \\
    & Emotion & 65.59 & 55.19 & 58.80 \\
    & Tandem & 72.87 & 53.53 & 59.83 \\
    \midrule
    \multirow{1}{*}{A3} & Multi-label & 71.95 & 52.59 & 58.74 \\
    \midrule
    \midrule
    & SeGA (Ours)& 68.13 & 56.58 & \textbf{60.69} \\
    \bottomrule
\end{tabular}
\end{adjustbox}

\end{table}

\subsection{Pre-Training Strategies Study}
To further validate the effectiveness of the pre-training design, we conducted extensive experiments with three aspects: prompt encoder (A1), prompt design (A2), and pre-training task (A2), as illustrated in Table \ref{tab:prompt}.

\subsubsection{A1: Variant of Prompt Encoder}
\begin{figure}
  \centering
  \includegraphics[width=\linewidth]{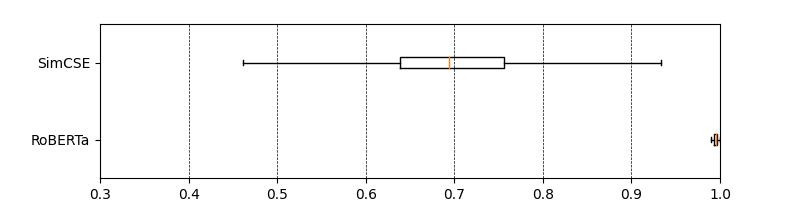}
  \caption{The first, second, and third quartiles of the cosine similarity results between prompt embeddings from RoBERTa and SimCSE.}
  \label{fig:boxplot}
\end{figure}

We changed the prompt encoder in Eq. (\ref{prompt-encoder}) from SimCSE RoBERTa to RoBERTa for preference-aware self-contrastive learning, which can be observed that replacing the prompt encoder with RoBERTa slightly decreases the performance.
As shown in Figure \ref{fig:boxplot}, we further compute the cosine similarity between prompt embeddings using these prompt encoders separately to analyze the discrepancies of different topic-emotion prompts.
The result shows that SimCSE RoBERTa provides more diverse discrepancies with different topic-emotion pairs in the same prompt template, while RoBERTa is inferior to separate different representations of topic-emotion pairs.
In addition, the minimum cosine similarity is 0.9795 with pre-trained RoBERTa and 0.3083 with SimCSE RoBERTa.
These results indicate that SimCSE RoBERTa is able to capture subtle differences between prompts arising from the replacement of emotions, whereas RoBERTa produces similar representations with different pseudo-labels.
The distinguishable embeddings enable performance enhancement for self-contrastive learning, which again raises the need for incorporating this for contrastive learning with prompts.

\subsubsection{A2: Variants of Prompt Design}
\label{prompt-variant}
To analyze different designs of the prompt template for self-contrastive learning, we evaluate SeGA with four variants: 1) \textbf{Short prompt}: Majority: $t^u_{i_{max}}$ - $e^u_{i_{max}}$, minority: $t^u_{i_{min}}$ - $e^u_{i_{min}}$.
2) \textbf{Topic prompt}: The majority of the posts express $t^u_{i_{max}}$, while a minority of them express $t^u_{i_{min}}$.
3) \textbf{Emotion prompt}: The majority of the posts express $e^u_{i_{max}}$, while a minority of them express $e^u_{i_{min}}$.
4) \textbf{Tandem prompt}: The majority of the posts express $t^u_{i_{max}}$, while a minority of them express $t^u_{i_{min}}$. The majority of the posts express $e^u_{i_{max}}$, while a minority of them express $e^u_{i_{min}}$.
For topic and emotion prompts, we calculate the corresponding preference frequencies independently, and the tandem prompt is an integration of topic and emotion prompts, which can be viewed as the prompt based on independent frequencies.

Adopting the short prompts for learning leads to an inferior performance compared with the proposed design, which implies that taking the semantics of natural language into account to form a prompt is essential for preference-aware self-contrastive learning compared to forming the template naively and structurally.
From topic and emotion prompts, the inclined effects indicate that considering the information of either topic of emotion to describe users is insufficient to differentiate normal, troll, and bot users.
Moreover, the performance decreases when applying tandem prompts, which illustrates that capturing the paired relationship between topics and emotions is more meaningful than considering them separately.
Meanwhile, our proposed prompt still outperforms the tandem prompt, which showcases the strength of not only jointly considering the frequencies, but also describing topic-emotion pairs together to form a prompt.

\subsubsection{A3: Variant of Pre-Training Task}
To delve into the learning strategies for incorporating topic-emotion pairs via posts, we modify the preference-aware self-contrastive learning to the multi-label classification task (Multi-label), which aims to predict all potential topic-emotion pairs of a user with the same model architecture.
We can see that the proposed self-contrastive learning outperforms predictive learning, which highlights that multi-label classification fails to capture users' preferred emotions associated with specific topics since it treats all labels as being equally important.
In contrast, preference-aware self-contrastive learning mitigates this limitation through the prompt design, leading to a substantial improvement in anomaly user detection.


\begin{figure}
    \centering
    \begin{subfigure}[t]{0.23\textwidth}
        \tikz\node[draw=black, inner sep=0pt] {\includegraphics[width=\textwidth]{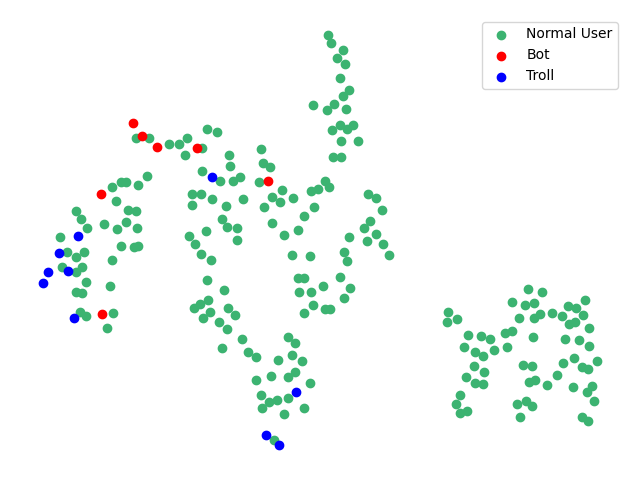}};
        \caption{SimpleHGN}
        \label{fig:img1}
    \end{subfigure}
    \hfill
    \begin{subfigure}[t]{0.23\textwidth}
        \tikz\node[draw=black, inner sep=0pt] {\includegraphics[width=\textwidth]{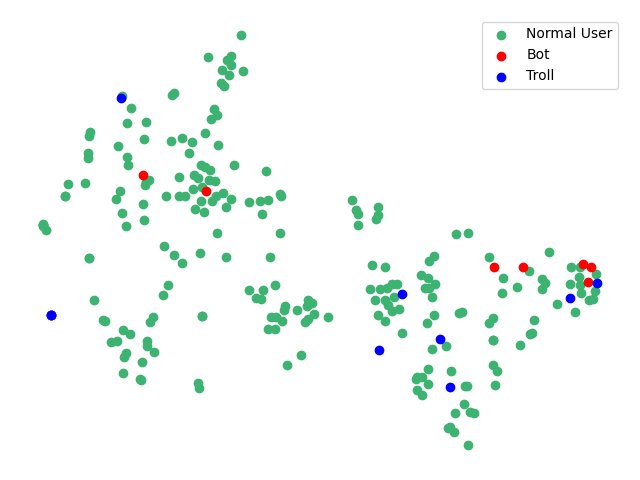}};
        \caption{SATAR}
        \label{fig:img2}
    \end{subfigure}
    \\
    \begin{subfigure}[b]{0.23\textwidth}
        \tikz\node[draw=black, inner sep=0pt] {\includegraphics[width=\textwidth]{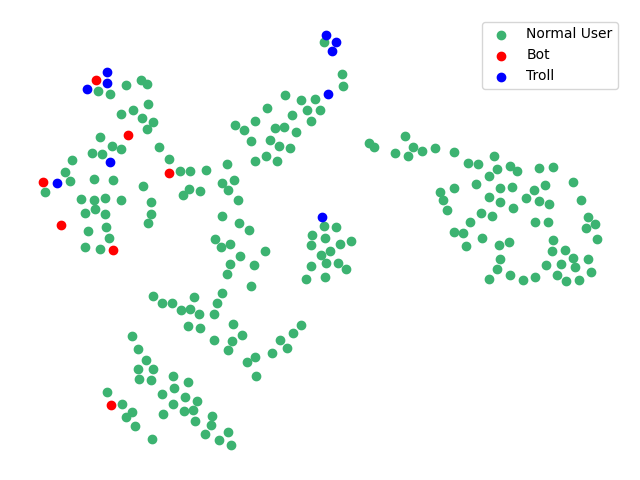}};
        \caption{RGT}
        \label{fig:img3}
    \end{subfigure}
    \hfill
    \begin{subfigure}[b]{0.23\textwidth}
        \tikz\node[draw=black, inner sep=0pt] {\includegraphics[width=\textwidth]{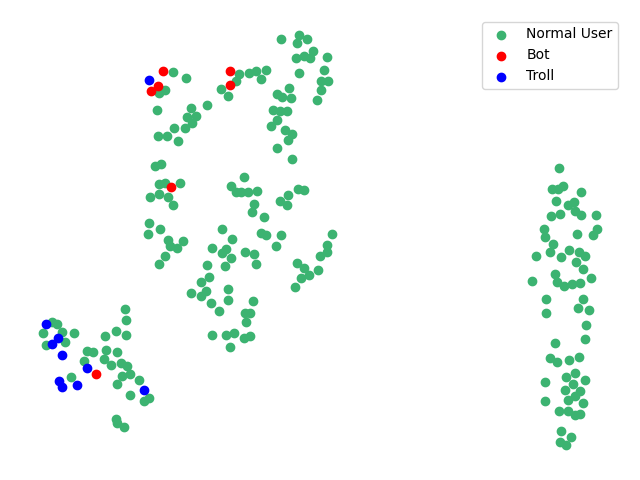}};
        \caption{SeGA (Ours)}
        \label{fig:img4}
    \end{subfigure}
    \caption{The visualization of user embeddings produced by the heterogeneous encoder and baselines.}
    \label{fig:embedding_visualization}
\end{figure}

\subsection{Case Study}
In order to investigate the efficacy of embeddings acquired from preference-aware self-contrastive learning related to anomaly user detection, we sampled users who primarily expressed anger in their general posts (i.e., others) to reflect the normal behavior of a user.
We note that extensive topics related to news are discussed in Appendix C.2.
We visualize the t-SNE embedding results of our approach against three baselines: RGT, SimpleHGN, and SATAR as illustrated in Figure \ref{fig:embedding_visualization}.
The embeddings of SATAR and RGT show ambiguous collocation for groups of normal, bot, and troll users, while troll and bot users of SimpleHGN have more distinguishable distances but mix with normal users.
Nonetheless, our proposed SeGA is able to describe different categories of users with clearer boundaries, which is attributed to preference-aware self-contrastive learning to describe user behaviors via the corresponding posts.
It is worth noting that the generic preference-aware objective is not directly intended to address the anomaly user detection task, but definite boundaries help group various user behaviors. 



\section{Conclusion}

This paper proposes SeGA, novel preference-aware self-contrastive learning with pseudo-preference generations for anomalous user detection on Twitter, including more challenging troll users.
Distinct from existing works that only focus on bot detection, our proposed method is able to distinguish various anomalous and normal users by learning similarities and discrepancies of topic-emotion pairs from posts of users summarized by LLMs, allowing the capability of capturing user-preferred behaviors.
Meanwhile, our prompt design considers the context of the multifaceted preferences of users to avoid the model being biased by only considering the most-appearance preferences.
We propose a new benchmark for distinguishing anomalous and normal users on Twitter, which shows that SeGA significantly outperforms state-of-the-art approaches by between 3.5\% and 27.6\%.
We believe that SeGA serves as a general framework for social media due to the flexible design for incorporating user preferences from external knowledge of LLMs as well as for the self-contrastive learning approach, and multiple interesting directions could be further explored within the framework, such as more metadata for user preferences, few-shot examples, etc.

\appendix
\section{Dataset Collection}

\begin{table}
\centering
    \caption{The instruction prompt and an example of the output of ChatGPT.}
    \label{tab:chatgpt}
    \newcommand{\specialcell}[2][c]{
  \begin{tabular}[#1]{@{}c@{}}#2\end{tabular}}


\begin{adjustbox}{center}
\begin{tabular}{p{5cm}|p{3cm}}
    \toprule
    Instruction prompt & Example output \\
    \midrule
    Please classify each tweet into the topics and corresponding emotions for the following ten posts. The available topics are arts \& culture, business \& finance, careers, entertainment, fashion \& beauty, food, gaming, hobbies \& interests, movies \& TV, music, news, outdoors, science, sports, technology, and travel. The emotions to consider are joy, sadness, anger, fear, trust, disgust, surprise, and anticipation. Please provide the classification for each post in the format 'topic - emotion'. Limit the response to less than 100 words. Following are the ten tweets numbered with '\#'. \textbackslash n & \begin{tabular}[t]{@{}p{5cm}@{}}
    1: news - anger\\
    2: news - anger\\
    3: news - anger\\
    4: news - anger\\
    5: news - anger\\
    6: news - anticipation\\
    7: news - anticipation\\
    8: news - joy\\
    9: news - anger\\
    10: news - anger
    \end{tabular} \\
    \bottomrule
\end{tabular}
\end{adjustbox}
\end{table}

We provide parts of the proposed TwBNT benchmark in the supplementary file due to the large-scale size of the dataset.

\subsection{Prompts for Acquiring Topic-Emotion Pairs from ChatGPT}
The instruction prompt P() in Eq. (11) and example output of topic-emotion pairs from ChatGPT are shown in Table \ref{tab:chatgpt}. 
This instruction prompt template is produced by ChatGPT by asking \textit{``If I want ChatGPT to classify 10 tweets into the topic and the corresponding emotion for each tweet, which prompt should I use?"}.
Afterwards, topic-emotion pairs are adopted for representing user preferences for preference-aware self-contrastive learning.

\subsection{Feature Descriptions}
\begin{table}
\centering
    \caption{Metadata of each entity.}
    \label{tab:metadata}
    \newcommand{\specialcell}[2][c]{
  \begin{tabular}[#1]{@{}c@{}}#2\end{tabular}}


\begin{tabular}{c|c}
    \toprule
    Entity & Metadata \\
    \midrule
    User & \specialcell[c]{created at, description, entities\\
    location, name, profile image url\\
    protected, url, username\\
    verified, withheld, followers count\\
    following count, tweet count, listed count}\\
    \midrule
    List & \specialcell[c]{private, created at, description\\
    name, follower count, member count}\\
    \midrule
\end{tabular}
\end{table}

\begin{table}
\centering
    \caption{The 16 categories of topics based on Twitter topics \cite{twitter_topic}, and 8 emotions based on Plutchik’s emotions \cite{DBLP:journals/corr/abs-1910-01340}.}
    \label{tab:topicemotion}
    \newcommand{\specialcell}[2][c]{
  \begin{tabular}[#1]{@{}c@{}}#2\end{tabular}}

\begin{tabular}{c|c}
    \toprule
    Type & Categories \\
    \midrule
    Topic & \specialcell[c]{ arts \& culture, business \& finance,\\
    careers, entertainment, fashion \& beauty,\\
    food, gaming, hobbies \& interests,\\
    movies \& TV, music, news, outdoors,\\
    science, sports, technology, travel}\\
    \midrule
    Emotion & \specialcell[c]{joy, sadness, anger, fear, trust,\\
disgust, surprise, anticipation}\\
    \midrule
\end{tabular}
\end{table}

The metadata in our proposed dataset TwBNT is shown in Table \ref{tab:metadata}.
In the node feature encoding stage, the numbers of indicator features k for the user and list node are 3 and 1, and the numbers of numerical features m for the user and list nodes are 5 and 4. 
For users, the 3 categorical features are: 1) profile image ownership, 2) protected status, and 3) verification status.
The 5 numerical features are: 1) account creation timestamp, 2) user name length, 3) the number of followers, 4) the number of followings, and 5) the number of tweets.
For lists, the categorical feature is privacy status, and 4 numerical features encompass: 1) the list creation timestamp, 2) list name length, 3) the number of followers, and 4) the number of members.
Table \ref{tab:topicemotion} summarizes the 16 topics and 8 emotions for LLMs to classify each tweet with the corresponding topic and emotion.
\section{Details and Setups of Baselines}
\subsection{Baselines}
The details of the baseline methods are depicted as follows:
\begin{itemize}
    \item GCN \cite{DBLP:conf/iclr/KipfW17} learns the user embedding by aggregating features from neighbor users equally.
    \item GAT \cite{DBLP:conf/iclr/VelickovicCCRLB18} aggregates features with the attention mechanism to model the different importance between users.
    \item SimpleHGN \cite{DBLP:conf/kdd/LvDLCFHZJDT21} adopts different strategies (e.g., learnable edge-type embeddings) to enhance GAT on the heterogeneous graph.
    \item SATAR \cite{DBLP:conf/cikm/FengWWLL21} jointly leverages tweets, metadata, and neighborhood information with a self-supervised representation learning framework to improve the user representation for bot detection.
    \item BotRGCN \cite{DBLP:conf/asunam/FengWWL21} constructs a heterogeneous graph and applies relational GCN \cite{DBLP:conf/esws/SchlichtkrullKB18} to capture diverse user relationships.
    \item RGT \cite{DBLP:conf/aaai/FengTLL22} proposes relational graph transformers to model the relation and influence heterogeneity between users and learn representations for bot detection.
\end{itemize}

\subsection{Implementation Details}
We initialize the node embeddings for GCN, GAT, and SimpleHGN with the same node feature encoding strategies as RGT.
All baselines follow the same hyper-parameter settings as the TwiBot-22 \cite{DBLP:conf/nips/FengTWWCZZZLYFZ22} evaluation framework.
\section{Addition Experiments}
\subsection{Hyper-Parameters Sensitivity}

\begin{table}
    \caption{The hyper-parameter sensitivity of SeGA.}
    \label{tab:hyper}
    \begin{tabular}{cc|cc|c}
    \toprule
    Type & Parameters & Precition & Recall & F1 \\
    \midrule
    SeGA                   &  & 68.13 & 56.58 & \textbf{60.69} \\ 
    \midrule
    \midrule
    \multirow{5}*{batch size}
    & 256                               & 72.11 & 51.96 & 58.08 \\
    & 512                              & 66.51 & 55.48 & 59.28 \\
    & 1024                              & 70.84 & 53.59 & 59.24 \\
    & \textbf{2048}                              & 68.13 & 56.58 & 60.69 \\
    & 4096                             & 70.04 & 52.39 & 58.12 \\
    \midrule
    \multirow{4}*{dropout}
    & 0.1                              & 65.68 & 55.58 & 59.34 \\
    & 0.2                              & 65.10 & 56.86 & 60.11 \\
    & \textbf{0.3}                              & 68.13 & 56.58 & 60.69 \\
    & 0.4                              & 70.39 & 51.61 & 57.14 \\
    \midrule
    \multirow{3}*{$d_{out}$}
    & 64                             & 75.03 & 50.91 & 57.40\\
    & \textbf{128}                            & 68.13 & 56.58 & 60.69 \\
    & 256                            & 70.97 & 53.86 &  59.40\\
    \midrule
    \multirow{4}*{$d_a$}
    & 16                               & 69.80 & 53.79 & 58.92\\
    & 32                               & 66.37 & 55.64 & 59.49 \\
    & \textbf{64}                               & 68.13 & 56.58 & 60.69 \\
    & 128                               & 65.76 & 52.67 & 57.23 \\
    \bottomrule
\end{tabular}
\end{table}

\begin{figure}
    \centering
    \includegraphics[width=\linewidth]{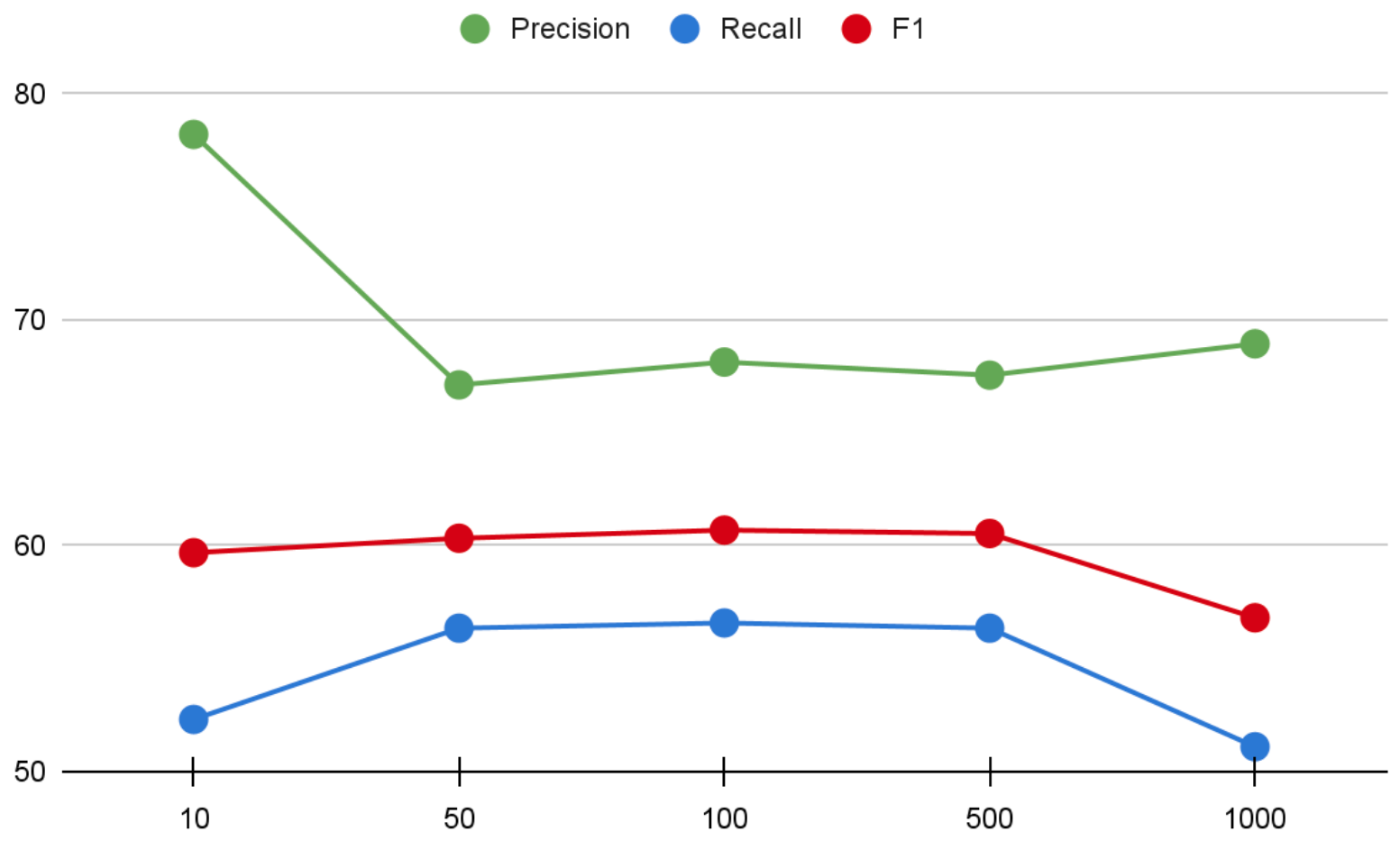}
    \caption{The effect of different numbers of negative samples.}
    \label{fig:negative}
\end{figure}


\begin{figure}
    \centering
    \begin{subfigure}{0.45\textwidth}
        \tikz\node[inner sep=0pt] {\includegraphics[width=\textwidth]{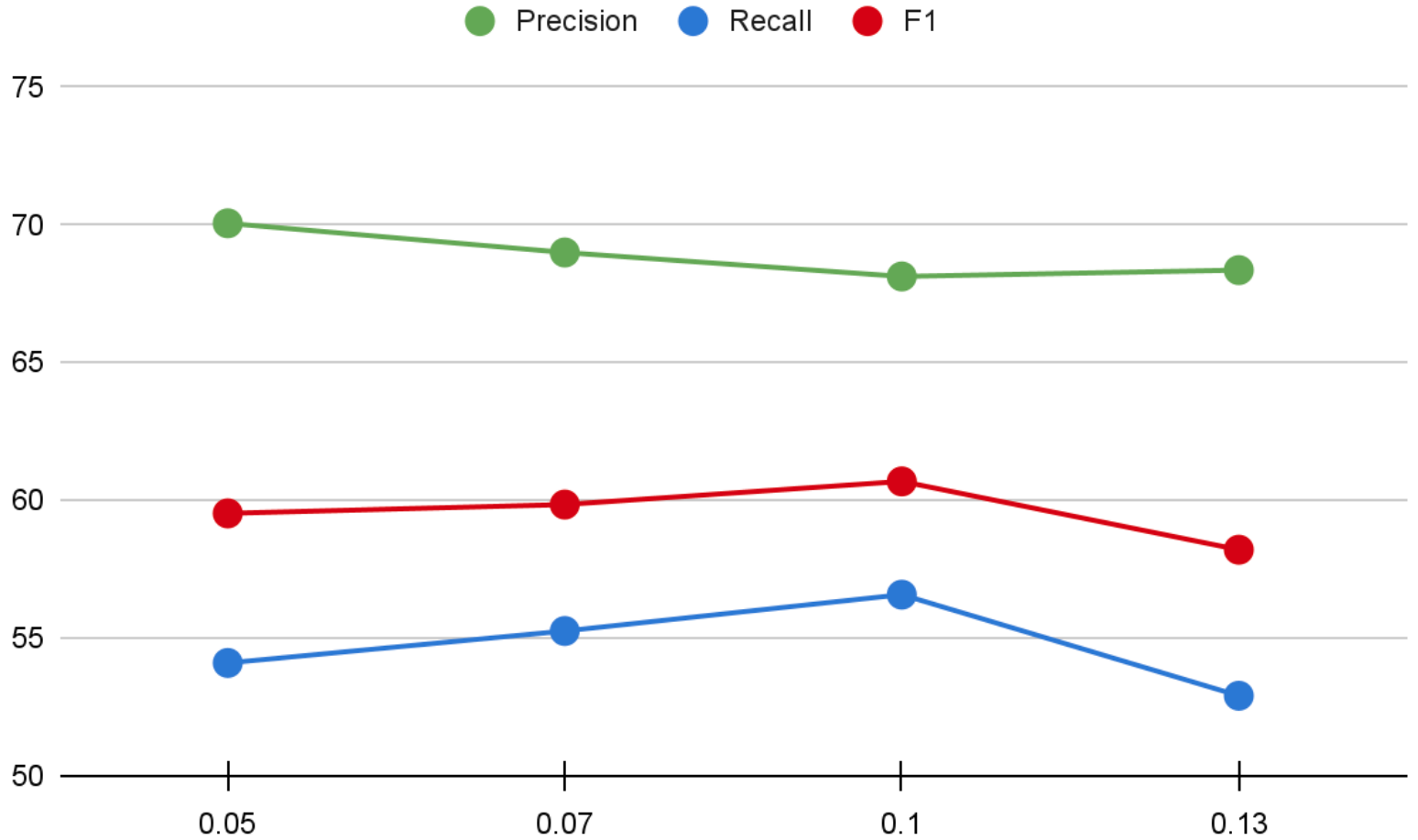}};
        \caption{Analysis of $\tau$ with \{0.05, 0.07, 0.10, 0.13\}.}
        \label{fig:t1}
    \end{subfigure}
    \hfill
    \\
    \begin{subfigure}{0.45\textwidth}
        \tikz\node[inner sep=0pt] {\includegraphics[width=\textwidth]{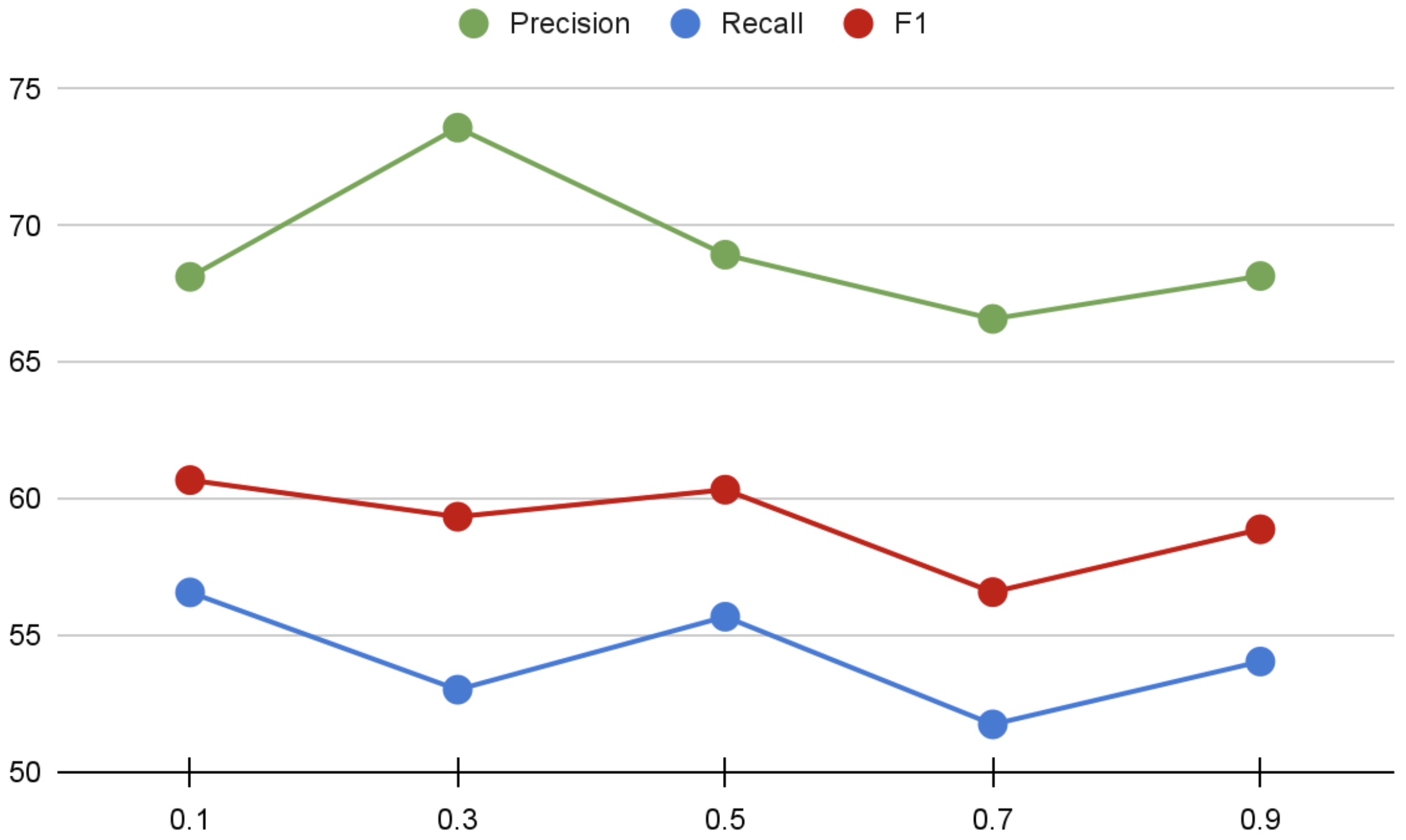}};
        \caption{Analysis of $\tau$ with \{0.1, 0.3, 0.5, 0.7, 0.9\}.}
        \label{fig:t2}
    \end{subfigure}
    \caption{The effects of the different numbers of temperature $\tau$.}
    \label{fig:temperature}
\end{figure}

To test the sensitivity of hyper-parameters, we report various settings of SeGA as shown in Table \ref{tab:hyper}.
The batch size is chosen from \{256, 512, 1024, 2048, 4096\}, and the dimension of embeddings $d_{out}$ and $d_a$ is chosen from \{16, 32, 64, 128\}  and \{64, 128,256\} respectively.
The dropout rate is chosen from \{0.1, 0.2, 0.3, 0.4\}.
It is observed that SeGA demonstrates robustness in terms of different hyper-parameter settings.

We further evaluate our model with different numbers of negative samples and temperatures $\tau$ for preference-aware self-contrastive learning.
Figure \ref{fig:negative} shows the performance of different numbers of negative samples with \{10, 50, 100, 500, 1000\}.
We observe a significant decrease in the F1 score when employing an excessively large number of negative samples, due to the model increased difficulty in converging for learning.
The impacts of different numbers of temperature $\tau$ are shown in Figure \ref{fig:temperature}, which can see that the performance of the F1 score is stable, which again illustrates the robust capability of our proposed framework.

\subsection{Extra Case Study}

\begin{figure}
    \centering
    \begin{subfigure}[t]{0.23\textwidth}
        \tikz\node[draw=black, inner sep=0pt] {\includegraphics[width=\textwidth]{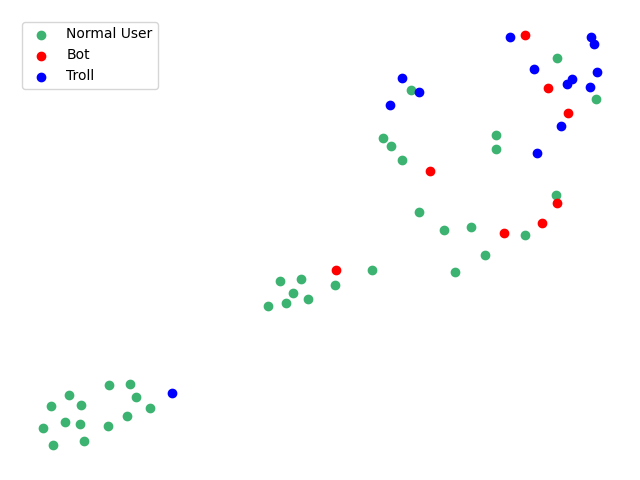}};
        \caption{SimpleHGN}
        \label{fig:img11}
    \end{subfigure}
    \hfill
    \begin{subfigure}[t]{0.23\textwidth}
        \tikz\node[draw=black, inner sep=0pt] {\includegraphics[width=\textwidth]{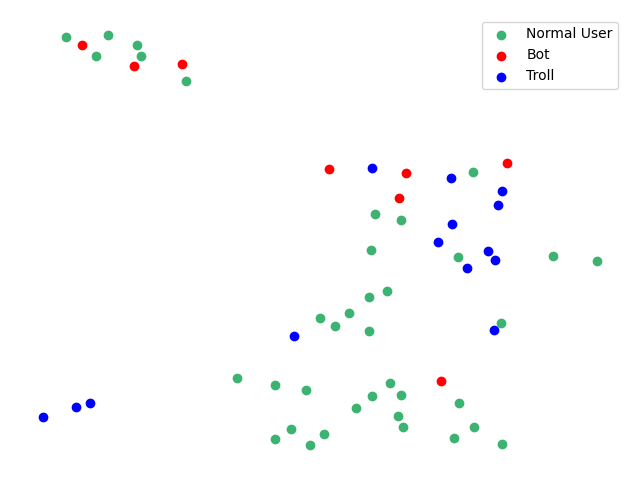}};
        \caption{SATAR}
        \label{fig:img22}
    \end{subfigure}
    \\
    \begin{subfigure}[b]{0.23\textwidth}
        \tikz\node[draw=black, inner sep=0pt] {\includegraphics[width=\textwidth]{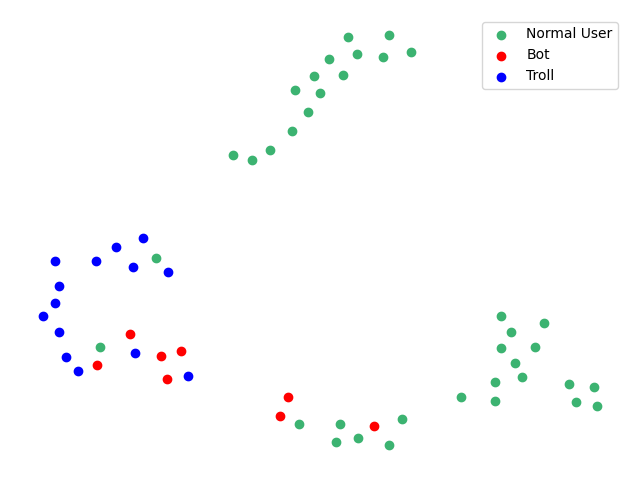}};
        \caption{RGT}
        \label{fig:img33}
    \end{subfigure}
    \hfill
    \begin{subfigure}[b]{0.23\textwidth}
        \tikz\node[draw=black, inner sep=0pt] {\includegraphics[width=\textwidth]{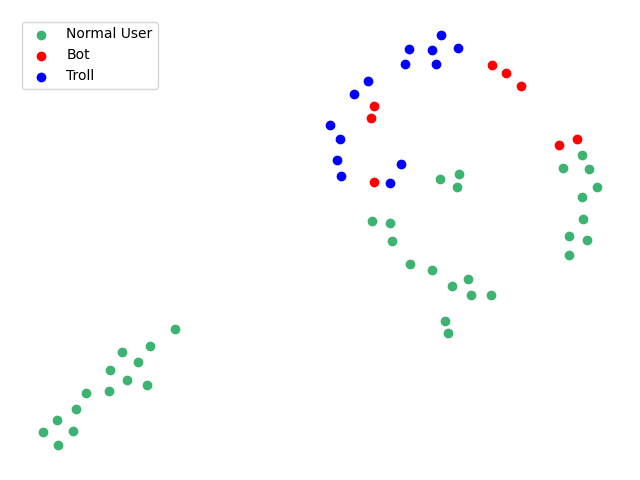}};
        \caption{SeGA (Ours)}
        \label{fig:img44}
    \end{subfigure}
    \caption{The visualization of user embeddings with different minor user preferences produced by the heterogeneous encoder and baselines.}
    \label{fig:case2}
\end{figure}

We further visualize the t-SNE embedding results with an extreme case shown in Figure \ref{fig:case2}.
We sampled users who predominantly post content related to angry news since topics related to news are often manipulated for malicious activities, and selected normal users who post minority topic-emotion pairs distinct from those of anomalous users.
Given this clear distinction, the model ideally should be capable of distinguishing between normal and anomalous users.
However, we can observe that existing methods (i.e., SimpleHGN, SATAR, and RGT) have struggled to effectively differentiate anomalous users.
In contrast, our proposed SeGA can effectively differentiate between anomalous users and normal users and establish a distinct boundary by leveraging these subtle differences in their preferences.

\bibliography{aaai24.bib}

\end{document}